\documentclass[prl,aps,twocolumn,groupedaddress,showpacs,floatfix]{revtex4}
\usepackage{amsmath,amssymb,multirow,epsfig,bm}

\newcommand{\beq}{\begin{equation}}
\newcommand{\eeq}{\end{equation}}
\newcommand{\bea}{\begin{eqnarray}}
\newcommand{\eea}{\end{eqnarray}}

\newcommand{\benn}{\begin{displaymath}}
\newcommand{\eenn}{\end{displaymath}}

\begin{document}

\title{Strongly paired fermions: Cold atoms and neutron matter}

\author{ Alexandros Gezerlis$^{1,2}$ and J. Carlson$^1$ }
\affiliation{$^1$ Theoretical Division, Los Alamos National Laboratory, Los Alamos, New Mexico 87545,  USA}
\affiliation{$^2$ Department of Physics, University of Illinois at Urbana-Champaign, Urbana, Illinois 61801, USA}

\begin{abstract}
Experiments with cold Fermi atoms can be tuned to probe strongly interacting
fluids that are very similar to the low-density neutron matter found in the crusts of neutron stars. 
In contrast to traditional superfluids and superconductors,  matter in this 
regime is very strongly paired, with gaps of the order of the Fermi energy.
We compute the $T=0$ equation of state and pairing gap for cold atoms and 
low-density neutron matter as a function of the Fermi momentum times
the scattering length.   Results of quantum Monte Carlo calculations 
show that the equations of state are very similar.
The neutron matter pairing gap at low densities is found to be very large but, 
except at the smallest densities, significantly
suppressed relative to cold atoms because of the finite effective range
in the neutron-neutron interaction.
\end{abstract}

\date{\today}


\maketitle

Strongly paired fermions are important in many contexts: 
cold Fermi atom experiments,  low-density neutron matter, and QCD at 
the very high baryon densities potentially found in the center of massive neutron stars.  
Developing a quantitative understanding of strongly paired Fermi systems
is important since they offer a unique regime for quantum many-body physics, 
relevant in very different physical settings including
the structure and cooling of neutron stars.
Constraining neutron matter properties can also be important in
understanding the exterior of neutron-rich nuclei by constraining
parameters of nuclear density functionals.

Cold-atom experiments can provide direct tests of the equation of state
and the pairing gap in the strongly paired regime, and hence
provide a crucial benchmark of many-body theories  in these systems.
We consider a system of  two fermion species and a simple Hamiltonian of the form
\begin{equation}
H = - \frac{\hbar^2}{2m} \sum_i \nabla_i^2 + \sum_{i<j} v(r_{ij}),
\end{equation}
where $i$ and $j$ represent spin up and down particles, respectively.
In cold atoms the interaction $v(r)$ can be tuned through Feshbach resonances
to be very attractive, and to produce a  specific scattering length.
By varying the scattering length one can sweep through various values
of Fermi momentum $k_F$ times scattering length  $a$ from
the BCS side (small $- k_F a$) to unitarity ($ - k_F a = \infty$) and beyond.
In many experiments $^6$Li atoms are used, and the effective range $r_e$
between the atoms is nearly zero. A variety of beautiful experiments
have been performed recently on cold atoms in the strongly paired
regime \cite{Joseph:2007,Bartenstein:2004,Zweirlein:2005,Luo:2007,Partridge:2006a,Partridge:2006b,Zwierlein:2006,Shin:2006,Shin:2007,Chin:2004}.

The neutron-neutron interaction, in contrast, is generally quite complicated,
with one-pion exchange at large distances, intermediate range spin-dependent
attraction dominated by two-pion exchange, and a short-range repulsion.  At very low densities, though, as found in neutron star
crusts or the exterior of neutron-rich nuclei, the scattering length
and effective range  are most crucial to the physical properties of the system. The presence
of a short-range repulsive core is important primarily in that it prevents collapse
to a higher-density state.

A strongly attractive zero-range interaction was proposed as a simple model
of neutron matter even before the recent remarkable cold atom experiments
\cite{Bertsch:MBX}. 
In low-density neutron matter the scattering length is
very large, $\approx -18.5$ fm, much larger than the typical separation between
neutron pairs.  The effective range is much smaller than the scattering length,
$r_e \approx 2.7$ fm, so $ |r_e / a| \approx 0.15$, but only at very low densities is the
effective range much smaller than the
interparticle spacing.

To the extent that the effects of finite range in the interaction can be neglected, cold atoms and neutron matter  are `universal' in the sense
that the properties of the system depend only upon the product of the Fermi momentum and the scattering length.
Experiments have been performed that probe the sound velocity \cite{Joseph:2007}  and collective excitations \cite{Bartenstein:2004}, superfluidity \cite{Zweirlein:2005} and critical temperature \cite{Luo:2007},  phase separation and phase diagram \cite{Partridge:2006a,Partridge:2006b, Zwierlein:2006,Shin:2006,Shin:2007}
and RF response \cite{Chin:2004}.

We have performed fixed-node quantum Monte Carlo (QMC) calculations for both cold atoms and neutron matter.  In each case, the trial wave function
is taken to be of the Jastrow-BCS form with fixed particle number and periodic boundary conditions:
\begin{equation}
\Psi_T = \  \left[ \prod_{i<j}  f (r_{ij}) \right] \  {\cal A} [ \prod \phi (r_{ij}) ].
\end{equation}
The BCS pairing function
$\phi (r)$ is parametrized with a short- and long-range part as in \cite{Carlson:2003}.
Since the ground-state energy in a fixed-node calculation is an upper bound to the true ground
state energy, we optimize the parameters to obtain
the lowest fixed-node ground-state energy, as in \cite{Carlson:2003}.

The interaction for cold atoms is taken as $v(r) =  - v_0 \frac{2 \hbar^2}{m} \frac{\mu^2}{\cosh^2(\mu r)}~$,
with $\mu = 24 / r_0$, or an effective range of $ r_0 / 12 $, with $1/ \rho = (4/3) \pi r_0^3$.  The interaction range is small enough not to significantly affect the
energy or pairing gap from the BCS regime to unitarity.
For the neutron-neutron interaction, we take the s-wave part of the
AV18 \cite{Wiringa:1995} interaction.  This interaction fits nucleon-nucleon scattering
very well at both low- and high-energies.  For our purposes the important thing
is that the scattering length and effective range are correctly described.
We use this interaction only between spin up-down pairs, which sets the
interaction in the $L=1, M= \pm 1$ pairs to zero. We correct for the artificial
attraction in the $L=1, M=0$ pairs perturbatively. This correction is $10 \%$
for the ground-state energy at the largest densities considered, and typically
much smaller.  The correction to the pairing gap is always smaller than the
statistical error in the calculation.

\begin{figure}
\begin{center}
\includegraphics[width=3.5in]{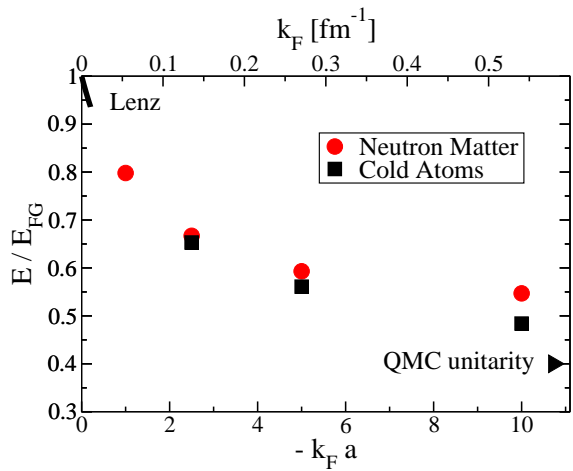}
\caption{Zero-temperature equation of state for cold atoms and neutron matter. Near zero density we show the analytical expansion of the
ground-state energy of a normal fluid, and at high density we show the cold atom
result at unitarity ($k_F a = \infty$, arrow). QMC calculations are shown as circles and
squares for neutron matter and cold atoms, respectively.}
\label{fig:eos-compare}
\end{center}
\end{figure}

The $T=0$ equations of state for cold atoms and neutron matter are compared in Fig. \ref{fig:eos-compare}.
The horizontal axis is $k_F a$, with the equivalent Fermi momentum $k_F$ for neutron matter shown
along the top.  The vertical axis is the ratio of the ground-state energy to the free Fermi gas energy $(E_{FG})$ at the
same density; it must go to one at very low densities and decrease as the density increases and
the interactions become important.
The curve at lower densities shows the analytical result \cite{Lenz:1929} for normal matter:
$ E/E_{FG}  =  1 + \frac{10}{9\pi}ak_F + \frac{4}{21\pi^2} \left
( 11 - 2 \ln2 \right ) \left ( ak_F \right )^2$.  This curve should be valid at very low densities. While it 
ignores the contributions of superfluidity, these are exponentially small in (1/$k_F a$).

The neutron matter and cold atom equations of state are very similar even for densities where the effective range is
comparable to the interparticle spacing.  Hence cold atom experiments can tell us something rather directly about the
neutron matter equation of state.  Near $k_F a = -10$ the energy per particle is not too far from QMC calculations \cite{Carlson:2003,Giorgini:2004}
and measurements \cite{Giorgini:2007} of the ratio $\xi$ of the unitary gas energy to $E_{FG}$; previous calculations give $\xi = 0.42(1)$.  Extrapolations of recent QMC calculations to $r_e = 0$ and also AFMC calculations suggest that $\xi = 0.40(1)$ \cite{Carlson:2007}  (arrow in Fig. \ref{fig:eos-compare}).

The results near $k_F a = -10$ for neutron matter are  compatible with previous calculations of the
neutron matter equation of state at somewhat higher densities ($k_F \geq 1$ fm$^{-1}$) \cite{Akmal:1998,CarlsonMorales:2003,Schwenk:2005}.
Results shown are for 66 particles in periodic boundary conditions;
calculations have also been performed near $N$ = 20, 44, and 90. Based on these results, finite-size effects for $N$ = 66 and beyond
are expected to be quite small, of the order of a couple percent.  Calculations of the cold atom equation of state
are very similar to those reported previously in \cite{Giorgini:2004} and \cite{Chang:2004}; the energies reported here are slightly lower (up to $\approx$ 10\% in
some cases) because of larger system sizes and better optimizations.

Realistic microscopic calculations that incorporate strong
pairing thus provide important constraints on the neutron matter equation
of state in the subnuclear saturation density regime.  Skyrme models
or more generally density functionals are used, for example, to determine
the structure of neutron star crusts \cite{Steiner:2005} and the neutron skin thickness of
nuclei \cite{skinthickness}.  A realistic treatment of these problems should incorporate 
the physics of the rapid transition of neutron matter from nearly free
particles to a strongly paired system at very low densities.

The pairing gap is the other fundamental zero-temperature property of
superfluid systems. Calculations of the s-wave pairing gap in neutron matter have
varied enormously over the past 20 years \cite{Lombardo:2001,Dean:2003}. The difficulties in accurately
calculating corrections to the BCS pairing gaps in the strongly paired regime
are significant,
and hence calculations of the pairing gap \cite{Lombardo:2001,Dean:2003,Chen:gap,Wambach:gap,Schulze:gap,Schwenk:gap,Fabrocini:gap} can differ by large factors
(from 4 to 10) in the low-density regime.  Cold atom experiments can provide
a critical test of theories of the pairing gap in this regime.  
We first compare our calculations of the pairing gap in cold atoms and neutron
matter, and then compare with previous results.

We calculate the pairing gap from the odd-even energy staggering $\Delta =
E(N+1) - ( E(N)+E(N+2))/2 $, where $N$ is an even number of particles.
Finite-size effects for the pairing gap are considerably larger than for the ground state energy.
In order to estimate the convergence of the gap to the continuum value with increasing $N$ we have
solved the BCS equations:
\begin{eqnarray}
\Delta({\bf k}) & = & -\sum_{{\bf k'}} \langle {\bf k} |V | {\bf k'} \rangle \frac{\Delta({\bf k'})}{2\sqrt{\epsilon({\bf k'})^2+\Delta({\bf k'})^2}} \nonumber \\
\langle N \rangle & = & \sum_{{\bf k}} \left [ 1 - \frac{\epsilon({\bf k})}{\sqrt{\epsilon({\bf k})^2+\Delta({\bf k})^2}} \right ]
\end{eqnarray}
in periodic boundary conditions for different $\langle N \rangle$.  

\begin{figure}
\begin{center}
\includegraphics[width=3.5in]{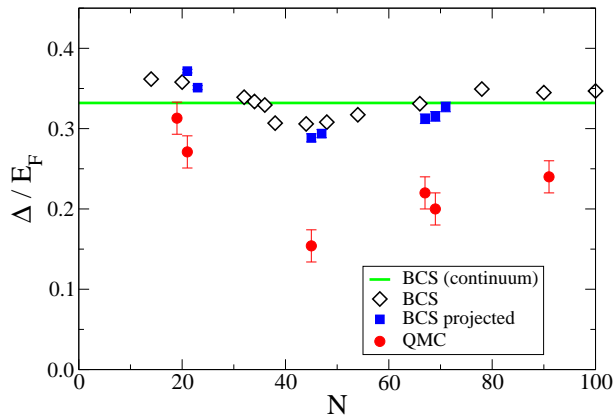}
\caption{Neutron matter pairing gap at $k_F a = - 10$ versus particle number in periodic boundary conditions, BCS and QMC calculations. }
\label{fig:gap-vs-n}
\end{center}
\end{figure}

The line in Fig. \ref{fig:gap-vs-n} is the continuum BCS result for $k_F a = -10$, and the open symbols
are the solutions of the BCS equations for different $\langle N \rangle$.
The continuum results are nearly
identical for the AV18 interaction and the simple cosh potential adjusted to yield
the same scattering length and effective range. For the finite systems BCS results are
shown for the cosh potential.
Unlike the case of cold atoms near unitarity, where $ - k_F a >> 1 $ and $r_e \approx 0$, the BCS gap
shows sizable oscillations for small numbers of particles.  The BCS value approaches the continuum limit (straight line) near $N=66$, and oscillations from that point on are fairly small, comparable in size to the
statistical error in the QMC calculations.   
We also show as solid points the gaps obtained from particle-projected
BCS wave functions in variational Monte Carlo calculations and the odd-even staggering formula. The projection to definite particle number is a small  effect.

\begin{figure}
\begin{center}
\includegraphics[width=3.5in]{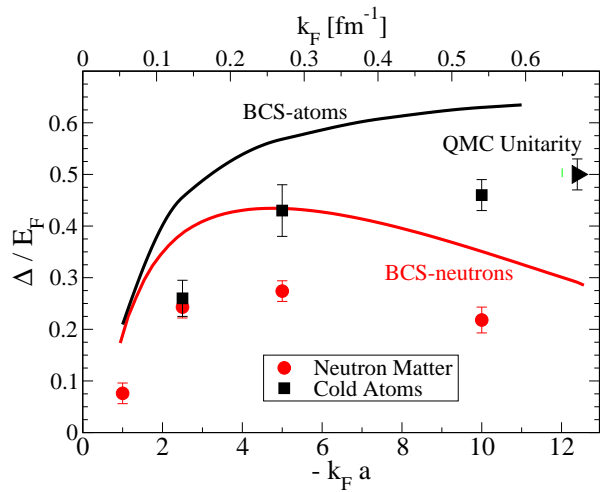}
\caption{Superfluid pairing gap versus $k_F a$ for cold atoms ($r_e \approx 0$) and neutron matter ($ |r_e / a| \approx 0.15$).  BCS (solid lines) and QMC results (points) are shown.}
\label{fig:gap-vs-kfa}
\end{center}
\end{figure}

The lower points in Fig. \ref{fig:gap-vs-n} are QMC results  for $k_F a = -10$.  
At very small values of N the gap  is quite large, as is also seen in the BCS calculations.   This is due to the  coarse description of the
Fermi surface in such small systems; the momentum grid spacing in occupied states is similar in magnitude to the Fermi momentum.
 When the pairing is very strong, as in 
cold atoms in the unitary regime, this coarse description is not too critical.
However for weaker coupling or  the larger effective range in neutron matter this becomes more 
important. The gap in both BCS and QMC calculations reaches a minimum near
44 particles (near the midpoint between closed shells at 38 and 54), and then increases to values near the continuum limit.
Pairing gap results for $N=66-92$ are consistent within the statistical errors.   

For all values of $N$ the gap is considerably smaller than the BCS results.  For comparison, at unitarity in cold atoms BCS calculations give a gap of $0.69$ $E_F$
while the QMC result is $0.50(3)$ $E_F$.\cite{Carlson:2005}  These calculations are
in good agreement with recent polarized cold atom experiments \cite{Shin:2007,Carlson:2007a}. For cold atoms the BCS equations will
produce the exact gap in the BEC limit where the pairs are strongly bound.
No such limit is relevant for a finite-range interaction.

In Fig. \ref{fig:gap-vs-kfa} we plot the pairing gap as a function of $k_F a$
for both cold atoms and neutron matter.   BCS calculations are shown as solid
lines, and QMC results are shown as points with error bars. QMC pairing gaps
are shown from calculations of $N=66-68$ particles.   For cold atoms away from
unitarity the pairing
gaps are smaller than calculated previously \cite{Chang:2004}, due to more complete optimizations 
and because these larger simulations reduce the finite-size effects.

For very weak coupling,
$ - k_F a << 1$, the pairing gap is expected to be reduced from the BCS
value by the polarization corrections calculated by Gorkov \cite{Gorkov:1961}
$\Delta/ \Delta_{BCS} = (1/4e)^{1/3}$.  Because of finite-size effects, it is difficult
to calculate pairing gaps using QMC in the weak coupling regime.
The QMC calculations at the lowest density, $k_F a = -1$, are roughly
consistent with this reduction from the BCS value.
At slightly larger yet still small densities,
where $-k_F a = {\cal O} (1)$ but $k_F r_e << 1$ for neutron matter, 
one would expect the pairing gap to be similar for cold atoms and neutron matter.  The results at $k_F a = -2.5$, where $k_F r_e \approx 0.35$, support this expectation.   Beyond that density the effective range becomes important and the QMC results are 
significantly reduced in relation to the cold atoms where $r_e \approx 0$.

\begin{figure}
\begin{center}
\includegraphics[width=3.5in]{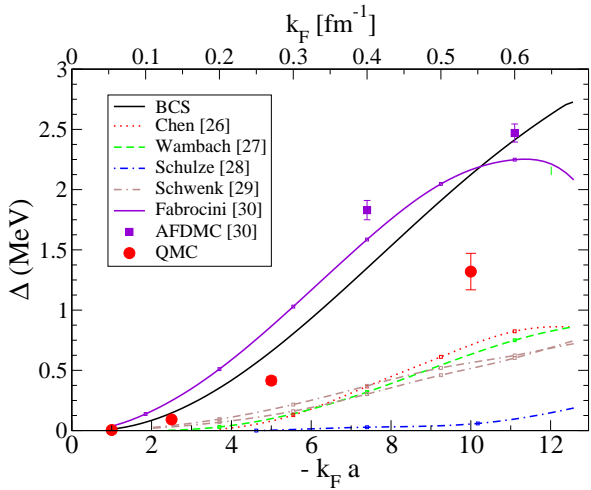}
\caption{Superfluid pairing gap versus $k_F a$ for neutron matter compared
to previous results.}
\label{fig:gap-compare}
\end{center}
\end{figure}

These results for the pairing gap are compared to selected previous results
in Fig. \ref{fig:gap-compare}. The results of our calculations are much
larger than the diagrammatic  \cite{Chen:gap,Wambach:gap,Schulze:gap} and renormalization group \cite{Schwenk:gap} approaches.
As these approaches assume a well-defined Fermi surface or calculate polarization
corrections based upon single-particle excitations it is not clear how well they can describe neutron matter in the strongly paired regime, or the
similar pairing found in cold atoms. 

The results here are significantly smaller than the AFDMC results of  Fabrocini {\it et al}.
\cite{Fabrocini:gap}.
These calculations are somewhat similar to those reported here.  The disadvantage
of the AFDMC approach is that it does not provide a variational bound
to the energy, and hence the wave functions are chosen from another approach.
In the calculation of Ref. \cite{Fabrocini:gap} the wave function was taken from a correlated basis
function approach that included a BCS initial state. The pairing
in that variational state is unusually large, and in fact increases as a fraction
of $E_F$ when the density is lowered. The advantage of the AFDMC approach is that it
includes the full interaction rather than the simple s-wave interaction used here.
AFDMC calculations with larger particle numbers are underway.\cite{Gandolfi:pc}

In summary, we have calculated the $T=0$ equations of state and pairing
gaps for cold atoms and neutron matter. These systems are quite similar
in that both are very strongly paired, and both have pairing gaps of the
order of the Fermi energy.  Experiments on the cold-atom equation of
state would be very valuable in constraining the neutron matter equation
of state.  Pairing gaps in neutron matter are found to be suppressed compared
to cold atoms and BCS theory, but much larger than in most other approaches.
Again, cold-atom experiments could provide very valuable tests of 
many-body theories. It could be very important to explore finite-range effects 
experimentally using other atomic systems.

\

{\it Note added in proof.}  We recently became aware of new calculations
of the equation of state and pairing gap for cold atoms using auxiliary
field Quantum Monte Carlo techniques \cite{Bulgac:2008}. Their results are similar to, but
slightly different than, those presented here.

\

A.G. wishes to express his gratitude to V. R. Pandharipande for initial guidance. The authors would also like to thank K. E. Schmidt for valuable discussions. 
The work of A.G. and J.C. is supported by the Nuclear Physics Office of the U.S. Department of Energy and by the LDRD program at Los Alamos National Laboratory.  The work on neutron matter is supported by the UNEDF SCIDAC program of the U.S. Department of Energy. Computing resources were provided at LANL and NERSC through the LANL Open Supercomputing Program and
SCIDAC. The work of A.G. was supported in part by NSF Grant Nos. PHY03-55014 and PHY05-00914.



\end{document}